\documentstyle[art8,aas2pp4,flushrt,tighten]{article}

\slugcomment{Astrophysical Journal, Vol. 551, April 20, 2001}

\lefthead{Heller \& Shlosman}
\righthead{ }

\begin{document}

\title{DOUBLE BARS IN DISK GALAXIES:\\DYNAMICAL DECOUPLING OF
NON-SELF-GRAVITATING GASEOUS BARS}

\author{Clayton Heller and Isaac Shlosman}



\affil{University of Kentucky, Lexington, KY 40506-0055, U.S.A. \\ email:
{\tt cheller@pa.uky.edu} and {\tt shlosman@pa.uky.edu}}

\and

\author{Peter Englmaier}

\affil{Max-Planck-Institut f\"ur Extraterrestrische Physik,
Karl-Schwarzschild-Strasse 1, Postfach 1312, Garching, Germany D-85741\\
email: {\tt ppe@mpe.mpg.de}} 

\begin{abstract}

We find that nuclear rings in barred galaxies can be subject to a new type of
non-self-gravitational dynamical instability. The instability leads to the
formation of {\it gaseous} molecular bars with pattern speeds which are
substantially {\it slower} than speeds of the primary {\it stellar}
bars. This spectacular decoupling of nuclear bars from the underlying 
gravitational potential is triggered but is not driven by the gas viscosity.
We find that low-viscosity systems can spend a substantial
period of time in a fully decoupled state, with the nuclear bar slowly
tumbling in the gravitational field of the primary bar. Higher viscosity
systems form nuclear bars which librate about the primary bar. The shape
of a nuclear bar, i.e., its eccentricity, correlates strongly with the angle
between the bars. We also find that such decoupling, partial or full,
most probably will be associated with bursts of star formation and with
gas inflow across the inner (Lindblad) resonance zone towards smaller
radii.  

\end{abstract}


\section{Introduction}

Still unknown, but a substantial fraction of disk galaxies exhibit a
phenomenon of double bars, stellar and gaseous (e.g., Shaw et al. 1995; Friedli
et al. 1996; Erwin \& Sparke 1999; Kenney 1997; Jogee 1999; Marquez et
al. 1999; Maciejewski \& Sparke 2000; Maiolino et al. 2000). The primary bars
extend for several kpc and in most cases define the galactic morphology. The
secondary (nuclear) bars, typically less than 1 kpc in extent, are observed
only at moderately high spatial resolution, either in the near-infrared or
through  the CO tracer of molecular gas. However, in order to study their
kinematics, an  HST-level of resolution is crucial. Both bar types are
expected to have a profound effect on the gas flow on corresponding spatial
scales and, therefore, on the galactic evolution. Gas is known to fuel
activity, stellar and nonstellar, but  nowhere else in a galaxy can it
influence the overall dynamics to such an extent as in the central regions. 
Gasdynamics in primary bars was investigated in great detail, including its
effects on the overall galactic evolution (e.g., Pfenniger 1996). Gas flow  in
the nuclear bars is largely assumed to be similar and a scaled-down version of
the flow in the large stellar bars (e.g., Regan \& Mulchaey 1999). The
validity of this approximation, however, is not based on theoretical analysis.

Despite the obvious importance of the nested bars configuration, our
understanding of its formation, dynamics and evolution is very limited.
Moreover, little work has been performed so far to address the specifics of
flow  in gaseous as compared to stellar bars. Even the sense of rotation of
nuclear bars and their pattern speeds have never been detected directly from
observations. 

Theoretically, three options exist. First, if the nuclear bar was formed via
self-gravitational instability (in the stellar or gaseous disks), it must spin
in the direction given by the angular momentum in the disk, i.e., in the
direction of the primary bar (Shlosman, Frank \& Begelman 1989). In this
case, the pattern speed of the nuclear bar, $\Omega_n$, will be substantially
{\it higher} than that of the primary bar, $\Omega_p$. This was confirmed in
numerical simulations which included the gas self-gravity (Friedli \& Martinet
1993; Combes 1994; Heller \& Shlosman 1994). The best arrangement corresponds
to the corotation radius of the nuclear bar being approximately equal to the
radius of the Inner Lindblad resonance (ILR) of the primary bar, reducing the
fraction of chaotic orbits (Pfenniger \& Norman 1990). The gas presence
appears to be imperative for this to occur  (Shlosman 1999 and refs. therein).
In this case, both bars are dynamically {\it decoupled} and the angle between
them in a face-on disk is arbitrary.

Second, two bars can co-rotate, being dynamically {\it coupled} and their
rotation completely synchronized. Such a configuration can be a precursor
to the future decoupled phase (discussed above), or continue indefinitely
(e.g., numerical simulations of Shaw et al. 1993; Knapen et al 1995). 
The nuclear bar is expected to lead the primary bar in the first quadrant
at a constant angle, close to $90^\circ$. An explanation for this phenomenon 
lies in the
existense of two main families of periodic orbits in barred galaxies, the
so-called $x_1$ orbits aligned with the primary bar and the $x_2$ orbits which
are perpendicular to it (e.g., Binney \&  Tremaine 1987). The gas losing
angular support due to the gravitational torques from the primary bar will
flow towards the center and encounter the region with the $x_2$ orbits, which
it will populate\footnote{Strictly speaking, the gas will not occupy the
perfectly periodic orbits because of dissipation, but will be found at nearby
energies, which is sufficient for our discussion}. The extent of the $x_2$
family defines the inner resonance region in the disk, i.e., position of the
nonlinear inner and outer ILRs. The nuclear bar may be further strengthened by
the gas gravity, which drags stars to $x_2$ orbits. However, the amount of gas 
accumulating in the ILR
resonance region may be insufficient to cause the dynamical runaway. In this
latter case, it is thought that the synchronized system of two nearly
perpendicular bars can be sustained indefinitely, until star formation or
other processes cease it.

Third, the secondary bar can rotate in the opposite
sense to the primary bar. This situation may arise from merging, when the
outer galactic disk acquires opposite angular momentum to that of the inner
disk. It was considered by Sellwood \& Merritt (1994) and Davies \& Hunter
(1997) and appeared as a non-recurrent configuration. At
least one of the bars should be purely stellar, because  the gas cannot
populate intersecting orbits. 

Although all three options discussed above have specific predictions verifiable 
observationally, the triggering mechanism(s) for the formation of such
systems require much better understanding. To close this gap, we analyze the 
decoupling
process of nuclear bars in disk galaxies. Here we show that
dynamical evolution does not stop with the formation of two coupled
perpendicular bars (as described above), {\it even} when gas self-gravity is
neglected. Instead, partial or complete decoupling of a nuclear gaseous
bar, depending on the degree of viscosity in the gas, is triggered for a
prolonged  period of time. This bar either librates around the major axis of
the primary bar, or acquires a different pattern speed, which is substantially
{\it slower} than that of the primary bar. Our work is complementary to  
studies which focused on large-scale gas dynamics in barred
galaxies, and, due to the limitations of the numerical schemes, did not resolve
the central regions discussed here (e.g., Athanassoula 1992).  The role of gas 
self-gravity in the
decoupling process is fully worked out in a subsequent paper, and in this
case $\Omega_n > \Omega_p$. Both regimes are likely to be relevant to
our understanding of galactic evolution.

Section~2 details the formation of the nuclear gaseous bar and its subsequent
decoupling from the primary bar. Section~3 discusses the physical reasons
for this evolution and possible observational implications for disk galaxies. 

\section{Nuclear Bar Evolution: Librations About the Primary Bar and Complete
Decoupling}

In our study, the disk and its bulge/halo are represented by Miyamoto \& Nagai
(1975) analytic potentials 

$$ \Phi = - {GM\over \sqrt{r^2+(A+B)^2}},  \eqno(1)  $$ 
where $M$ --- mass in units of $10^{11}~{\rm M_\odot}$ and $A, B$ --- spatial
scaling parameters, in units of 10~kpc, are given in Table~1. 

\begin{table}
\caption{Model Parameters (see text)}
\smallskip
\begin{tabular}{lccc} \hline\hline
Component & Mass & A & B \\ \hline
disk      & 1.0  & 0.4  & 0.1  \\
bulge     & 0.3  & 0.0  & 0.1  \\ \hline
\end{tabular}
\end{table}    

The large-scale stellar bar is given by the Ferrers' (1877) potential with
$n=1$ and rotates with a prescribed pattern speed $\Omega_p$. In dimensionless
units, its mass and semi-major ($a$) and semi-minor ($b$) axes are 0.18, 0.22
and 0.05, respectively. The dynamical time $\tau_{\rm dyn}=1$ corresponds to
$4.7\times 10^7$~yrs.   The gas is evolved using a 2D version of a Smooth
Particle Hydrodynamics (SPH) code (details in Heller \& Shlosman 1994)
neglecting the gas self-gravity, and, alternatively, using the grid code
ZEUS-2D (Stone \& Norman 1992). 

The bar potential is turned on gradually to
avoid transients. The pattern speed  $\Omega_p=1.1$ and the bar strength are
chosen to form a double ILR in the disk and allow the primary bar to end near
the Ultra-Harmonic resonance at $r=0.7$, with corotation at $r=0.9$. Formation
of the ILRs is verified using nonlinear orbit analysis (see Heller \& Shlosman
1996). Along the primary bar major axis the outer ILR is found at $r=0.14$ and
inner ILR at 0.05. Note, that due to the primary bar strength, linear analysis
gives erroneous results about positions of the ILRs. Gas sound speed is taken 
to be $10~{\rm km\ s^{-1}}$.

\figurenum{1ab}
\begin{figure}[ht!!!!!!!!!!!!!]
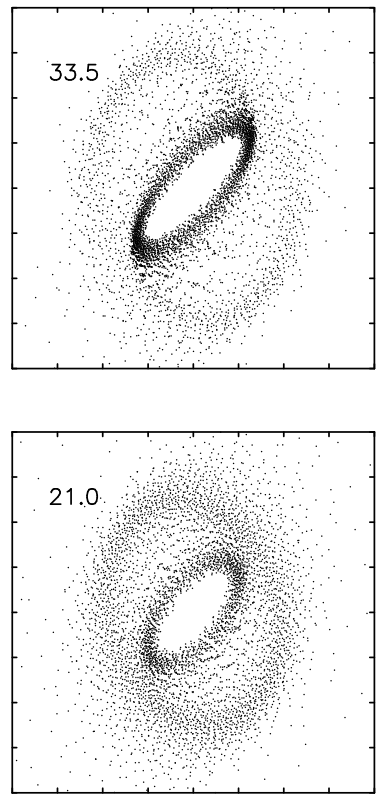

\vbox to4.6in{\rule{0pt}{4.6in}}
\includegraphics{fig1a.ps}
\includegraphics{fig1b.ps}
\caption{Representative frame in the evolution of ($a$) the Standard model
(upper) (Seq.~1) and ($b$) high-viscosity model (lower) (Seq.~2). See
Fig.~1c captions and the text for future details. The animation sequences
discussed here are available in the online edition of the Journal.}  
\end{figure}
\figurenum{1c}
\begin{figure*}[ht!!!!!!!!!!!!!!!!!]
\vbox to7.25in{\rule{0pt}{7.25in}}
\includegraphics{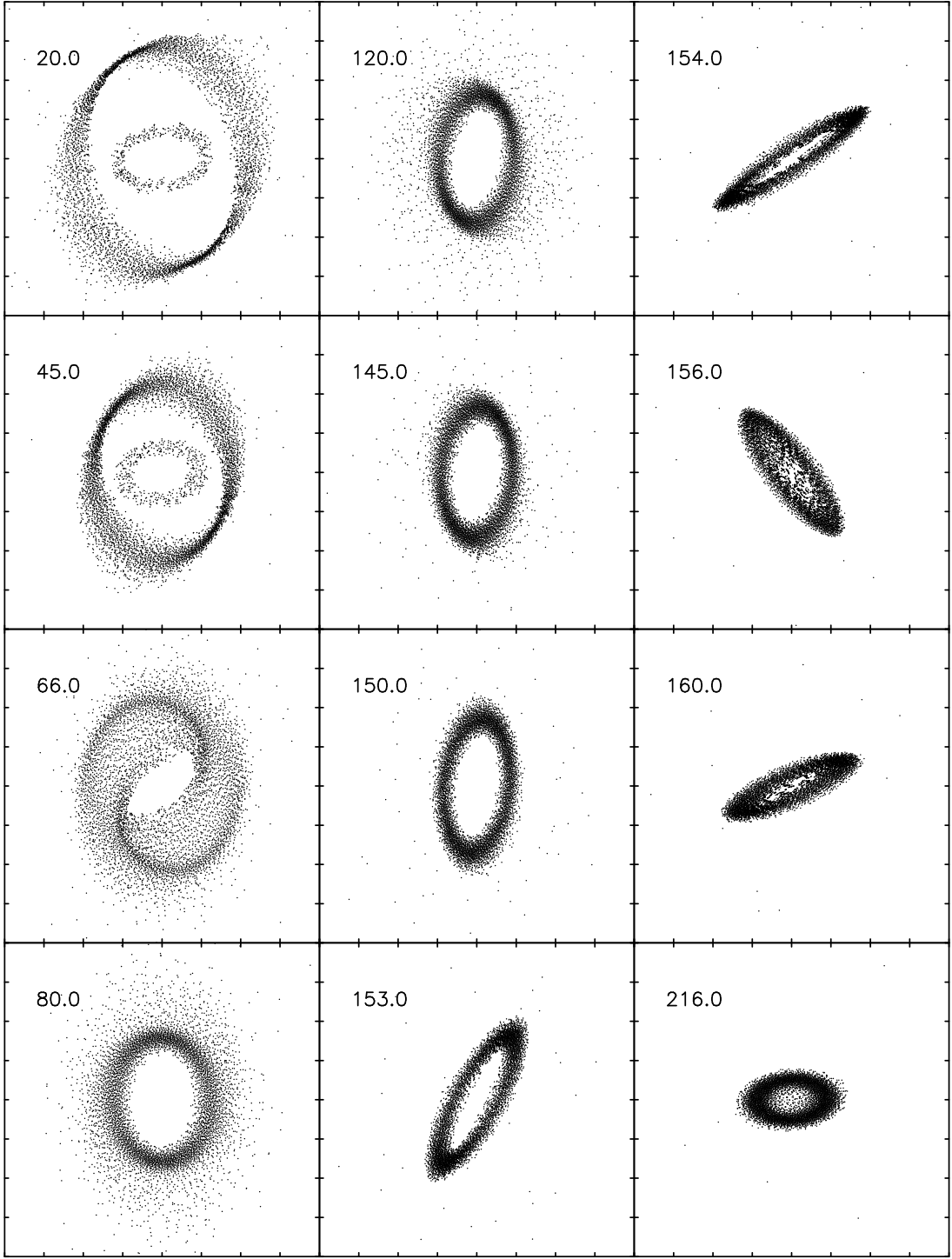}
\caption{Time evolution of the low-viscosity model (Seq.~3): 2D SPH simulation
in the background gravitational potential of a barred disk galaxy (shown
face-on). The gas response to the bar torquing is shown in the primary bar
frame. The primary bar is horizontal and the gas rotation is
counter-clockwise. Note a fast evolution after $t\sim 150$, when the nuclear
bar decouples and swings clockwise (!). The bar is `captured' again at $t\sim
211$. Time is given in units of dynamical time $\tau_{dyn}$. The animation
sequences discussed here are available in the online edition of the Journal.}
\end{figure*}      

The evolution of our Standard model is shown in the animated Sequence~1 in
the frame of the primary bar. We ran two additional models with identical 
initial conditions to that in the Standard model. The {\it only} difference is
that the value of viscosity was changed by a factor of 2, up and down (hereafter 
`high'
[Seq.~2] and `low' [Seq.~3] viscosity models). This was achieved by varying
the coefficients of viscosity $\alpha$ and $\beta$ in the prescription given
by Hernquist \& Katz (1989). As expected, the gas responds to the
gravitational torque from the bar by forming a pair of large-scale shocks,
loses its angular momentum and accumulates in a double ring (corresponding
roughly to the inner and outer ILRs, see below). In Seqs.~1 (also Fig.~1a) and
2 (also Fig.~1b), both rings interact hydrodynamically and as a result, the
outer ring is destroyed and its gas flows into the inner (nuclear) ring. In
the low viscosity model (Seq.~3 and Fig.~1c),
the inner ring is destroyed and its gas is mixed with the outer ring at
intermediate radii. After merging, a single oval-shaped ring corotates with
the primary bar, leading it by $\phi_{\rm dec}\sim 50^\circ$ (hereafter, the
angle at decoupling) for the high-viscosity Seqs.~1 and 2, and by
$\sim85^\circ$ for the low-viscosity Seq.~3. The angle $\phi_{\rm dec}$ while
depending on the gas viscosity, is always located in the first quadrant.

\figurenum{2ab}
\begin{figure}[ht!!!!]
\vbox to2.4in{\rule{0pt}{2.4in}}
\includegraphics{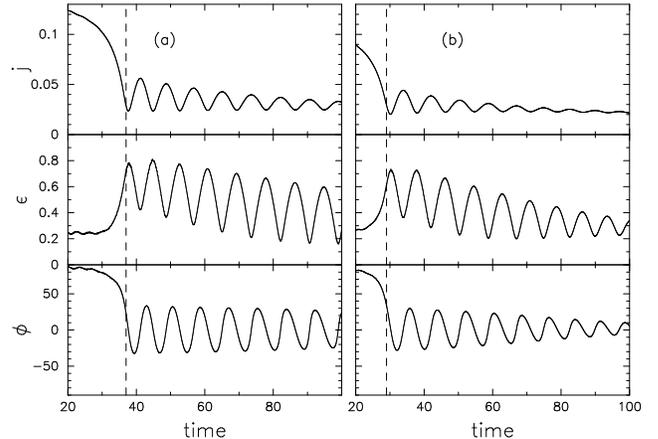}
\caption{Standard ($a$) and high viscosity ($b$) models: evolution of total
specific angular momentum $j$ in the primary bar frame of reference (upper),
eccentricity $\epsilon = 1-b/a$ (middle) and position angle $\phi$ of the line
of apses (lower) of the gaseous nuclear bar. Both $\phi$ and $\epsilon$ were
computed from the tensor of inertia of gas within $r\leq 0.2$. The time is
give in units of $\tau_{\rm dyn}$. The vertical dashed lines show the end of
the merging process between the rings.}
\end{figure}         

Next, the remaining ring becomes increasingly oval and barlike (Figs.~2a,b),
its pattern speed changes abruptly, and it swings towards the primary
bar, {\it against} the direction of rotation of this bar. We describe
the evolution of this gas flow in terms of the bar dynamics and call it a
nuclear bar. (In the inertial frame, the nuclear bar still spins in the same
direction as the main bar, albeit with a smaller pattern speed than
$\Omega_p$.) The shape of this bar can be described by its axial
(minor-to-major) ratio which reaches a minimum when both bars are aligned.
After crossing the primary bar axis into the 4th quadrant, the nuclear bar
initially decreases its axial ratio, i.e., becomes less oval. Thereafter
it stops, becomes more oval and speed up in the prograde direction. The bar
axial ratio is reaching its minimum when both bars are aligned and the gaseous
bar rotates in the retrograde direction (in the primary bar frame!). Hence,
in Seqs.~1 and 2, the nuclear bar librates about the primary bar with a
decreasing amplitude, being damped more strongly in the higher-viscosity
model. The low-viscosity Seq.~3, however, behaves in a qualitatively different
way. Instead of librating about the primary bar, the nuclear bar continues to
swing in the same direction maintaining a pattern speed $\Omega_n < \Omega_p$
for about 60 dynamical times, corresponding in our units to about $2-3\times
10^9$~yrs (Figs.~1c, 2c). On average, its pattern speed is about half of
$\Omega_p$, oscillating around this value with a substantial amplitude. At
times, $\Omega_n$ is very small, giving the impression that the nuclear bar
stagnates in the inertial frame of reference. With time, the
eccentricity of the nuclear bar gradually decreases and the bar is trapped
again by the valley of the potential of the main bar, thus entering the
libration phase, similar to Seqs.~1 and 2.

\figurenum{2c}
\begin{figure}[ht!!!]
\vbox to2.7in{\rule{0pt}{2.7in}}
\includegraphics{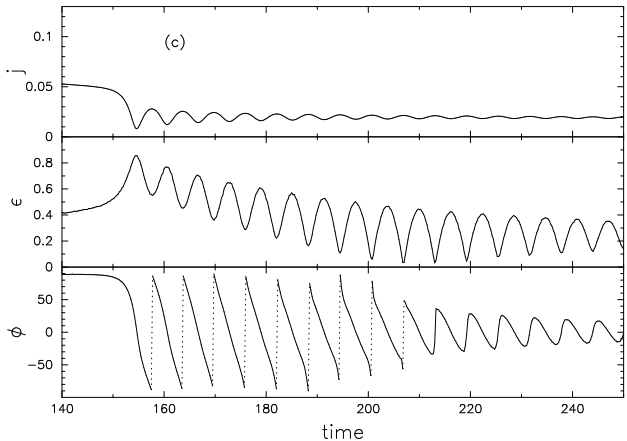}
\caption{Low-viscosity model: as in Figs.~1a,b. Angle $\phi$ is shown
between $\pm\pi/2$ for presentation only and discontinuities are marked
by vertical dotted lines. Full decoupling of the nuclear bar exists
at $\tau\approx 150-211$.}
\end{figure}   

This behavior of a nuclear gaseous bar, its formation and librations, and
especially the kinematically decoupled phase of evolution was never
previously observed in numerical simulations. In the following we provide
a theoretical explanation for this phenomena within the framework of orbital
dynamics in disk galaxies.

\section{Discussion: Non-Self-Gravitating Nuclear Gaseous Bars as Slow Rotators}

In this paper we neglect the self-gravitating effects in the gas. This
situation can be encountered in the early type and/or gas deficient barred
disks. A large-scale bar in such a 
galaxy torques the gas, which falls towards the central kpc accumulating in
the vicinity of the ILR(s) in the form of a nuclear ring. If the gas surface
density and the overall gas accumulation in the ring are small enough, both
local and global self-gravitational instabilities will be supressed. This is
the situation we envision. Note, that the positions of ILRs are
model-dependent and reflect the overall mass distribution in a galaxy. 
If the ILRs are absent initially, this will not affect the instability
discussed here, as they will form quickly in response to a growing central mass
concentration (e.g., Hasan \& Norman 1990). The evolution in all models was
similar in the SPH and ZEUS-2D codes, hence only the former is shown.

The high resolution and our emphasis on the gas kinematics in the
inner resonance region of disk galaxies allow us to study the evolution of
nuclear gas with different viscosities. All models, after some intermediate
evolution lead to the formation of a single ring, which quickly acquires a
barlike  shape with well-defined major and minor axes, and librates or
precesses rigidly with a pattern speed independent of radius (although the
shape is constantly varied). We, therefore, treat this configuration as a
bar in all respects. Gaseous bars differ of course from the stellar bars in
that the gas can only reside on nonintersecting and/or not excessively curved
orbits. Otherwise, shocks  develop and the gas quickly depopulates these
orbits. This difference extends also into non-self-gravitating regime.

The most dramatic behavior of gaseous bars in our models is observed in the
least viscous system. Here, in the primary bar frame, the nuclear bar decouples
and swings in the retrograde direction until being captured again by the
primary potential valley. In the inertial system, it spins in the prograde
direction, but its $\Omega_n$ is much smaller than $\Omega_p$. The precession
rate of the gaseous bar is the averaged precession rate of orbits within it,
which depends both on the underlying potential {\it and} the shape of the
orbit. Models with a higher viscosity show nuclear bar relaxation towards the
primary bar in the form of damped librations. 

Why is the gaseous bar precessing as a solid object despite the absence of
self-gravity? A simple exercise, of replacing the SPH gas by 
collisionless `stellar' particles, shows that in such a case each particle
orbit precesses at a slightly different rate and the bar dissolves into an
axisymmetric configuration in a few orbital times. Hence, a corresponding {\it
stellar} bar without self-gravity cannot exist. But in the limit of
weak gravity, it will have a confining effect on stars as well as on
gas, and our conclusions will still hold. Addressed elsewhere will be the case
of a dominant gravity, when self-gravitational modes are excited with pattern
speeds well above $\Omega_p$ and overtake the other modes.

It is clear, therefore, that it is the `fluid' (i.e., collisional) nature of
the gas which acts to preserve the bar despite the presence of internal shear
whose effect is to widen the flow. This process does {\it not} lead to
circularization of the orbits because the background potential does not admit
circular orbits. Dependence of bar eccentricity on radius
also contributes to the shear, which manifests itself in the gradual azimuthal
shift of lines of apses for individual orbits in the bar.  At other azimuths,
however, the orbits are re-focused again, preventing the bar from dissolving.
Because the gas is not self-gravitating in our approximation, the gravity
cannot cause the apse alignment and rigid precession of the ring, using the
mechanism of Godreich \& Tremaine (1979). Instead,
this re-focusing is purely hydrodynamical in nature and involves the action of
oblique shocks (hydrodynamical torques, i.e., bulk viscosity) in the flow.   Of
course, such confinement  results in the loss of energy by the flow, which is
radiated away by the isothermal gas in the simulations. This loss of energy by
the gas leads to non-conservation of Jacobi energy $E_J$ along the
orbit.\footnote{Jacobi energy is a {\it conserved} total energy of a gas
parcel in the rotating frame of the primary bar when dissipation is neglected 
(see e.g.,  Binney \& Tremaine 1987 for more details).} Before the rings merge,
this effect is largest on the outer ring because it forms at energies where
the shear is larger. After merging, the remaining ring slowly migrates to
lower energies until it encounters the inner ILR, becomes barlike and
decouples.

Next we address the physical reasons for the partial or complete runaways shown
in each model. The concept of periodic orbits in barred potentials is
central to our understanding of both the stellar and gas dynamical evolution in
these systems. Most importantly, the families of periodic orbits 
$x_1$ and $x_2$, mentioned in section~1, are oriented differently with respect
to the large-scale bar. Specifically, the $x_1$ orbits are aligned with the
bar between the outer ILR and the corotation radius and between the center and
the inner ILR. The $x_2$ orbits are perpendicular to the bar between the ILRs.
The existense of $x_2$ orbits is a result of the ``donkey'' effect
(Lynden-Bell 1979). A typical orbit will precess with respect to the bar due
to the gravitational torques from the latter. These orbits can be `captured'
by the bar potential when the precession rate is sufficiently close to
$\Omega_p$. Trapped orbits will librate either about the valley of the
potential ($x_1$ orbits), i.e. about the bar major axis, or at right angles to
the primary bar ($x_2$ orbits). As shown by Lynden-Bell, some orbits will
speed-up when pulled back by the bar (the ``donkey'') and some will slow down
in response. The former orbits will have a stable orientation along the minor
axis of the bar and contribute to the $x_2$ family, while the latter will
orient themselves along the major axis and contribute to the $x_1$ family. 
It is worth reminding that the $x_1$ and $x_2$ orbits are resonant orbits and
precess with the primary bar pattern speed $\Omega_p$. So in the primary bar
frame they do not change their orientation.

The extent of the $x_2$ orbits can only be determined self-consistently from
a nonlinear orbit analysis. Results are presented in Fig.~3 which
displays the characteristic curve of the $x_2$ family for the full range in
$E_J$. For the potential at hand, $E_J$ extends from $-4.6$ to $-2.7$ in our
units. It is the lower limit in $E_J$ of the $x_2$ orbits which is important,
as evident from the $E_J$ distribution of the SPH particles in the nuclear bar
located approximately at the inner ILR at the time of decoupling $\tau=150$
(Fig.~4). The exact value of this limit is model-dependent of course, but this
is of no importance to the essence of the decoupling.  

\figurenum{3}
\begin{figure}[ht]
\vbox to2.6in{\rule{0pt}{2.6in}}
\includegraphics{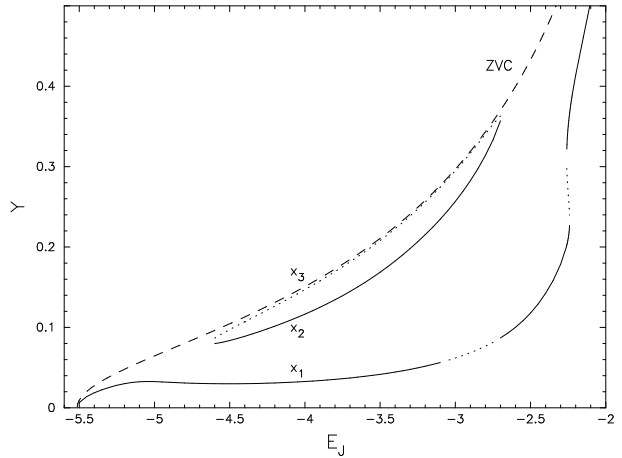}
\caption{Characteristic diagram for the potential and $\Omega_p$ used here.
The $y$-axis gives the orbit intercepts along the minor axis of the primary
bar. Only $x_1$, $x_2$ and $x_3$ families are shown. The dotted lines
represent unstable orbits. The dashed  curve labeled ZVC is the zero
velocity curve. The ILRs are located at $E_J=-4.6$ and $-2.7$.}
\end{figure}
\figurenum{4}
\begin{figure}[ht!!!!!!!]
\vbox to4.5in{\rule{0pt}{4.5in}}
\includegraphics{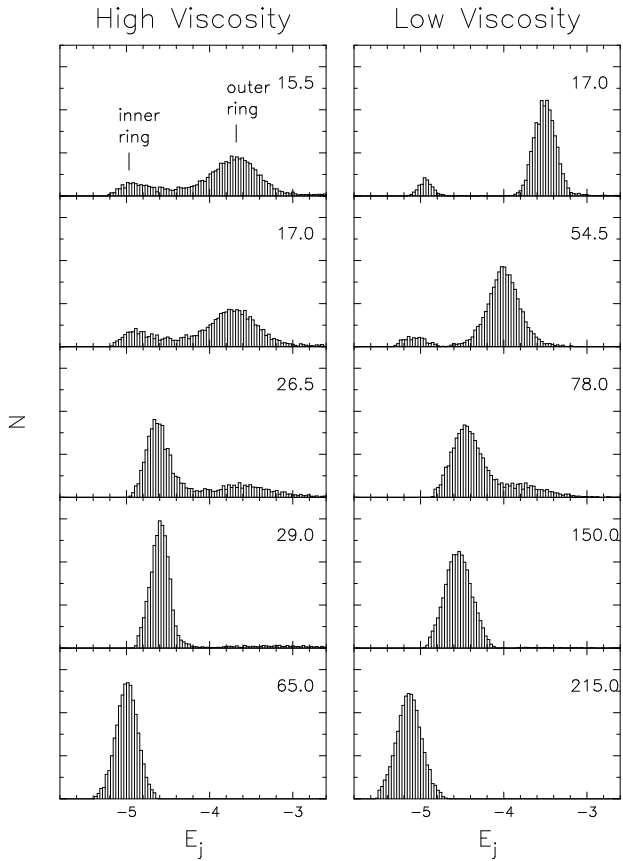}
\caption{Evolution of SPH particle distribution with the Jacobi energy $E_J$
for the high-viscosity (left) and low-viscosity (right) models.
The sequence depicts the times of formation of two rings, their pre- and
post-merging configurations, decoupling and post-decoupling of nuclear
bar. Note that on the left, merging is accomplished at $\tau\approx 29$ {\it
after} the bar enters libration phase at $\tau\approx 26.5$. The $x_2$ orbits
exist only between $E_J=-4.6$ and $-2.7$ (Fig.~3).}
\end{figure}                        

As seen in Figs.~1c, 3 and 4, and Seqs.~1-3, the inner ring forms just inside 
the inner ILR (in $E_J$ and $r$), on the $x_1$ orbits and is aligned with the
primary bar. Its gas was swept up from orbits within the inner
ILR by a positive torque from the primary bar. The outer ring forms between
the ILRs, on $x_2$ orbits of intermediate $E_J$, and is aligned
with the minor axis of the bar. Viscosity is most important during
the phase when both rings interact hydrodynamically and merge, which
understandably starts earlier and proceeds faster for higher viscosity models.
This pre-decoupling evolution is similar in all models, apart from the
timescale. 

The crucial difference between the models comes from (1) position
of the nuclear bar on the $E_J$ axis after the merging, and (2) the value of
$\phi_{\rm dec}$ angle, i.e. the orientation of the nuclear bar at the time of
decoupling. For high-viscosity and standard models the gaseous bar is situated
very close to the inner ILR at $E_J=-4.6$, with a dispersion of
$\pm0.1$. The low-viscosity bar resides at $-4.5$ and is accompanied by an
asymmetric `high'-energy tail which extends to about $E_J=-3$ (Fig.~4) 
representing a gaseous `envelope' around the bar (see Fig.~4 and Seq.~3).
We note that the role of viscosity here is fundamentally different from that
in nearly axisymmetric potentials (planetary rings, galactic warps), where it
acts to circularize the orbits. In barred potentials, circular orbits do not
exist and the gas merely moves from one elongated orbit to another.

The evolution of the inner ring consists of a few phases: formation, merging
with the outer ring, migration to the inner ILR, decoupling (partial or
full) and post-decoupling. The first phase was discussed above, and the
later evolution is addressed below. 

The merging with the outer ring mixes high and low-$E_J$ gas and the ring is
`lifted' across the inner ILR, acquiring a leading angle $\phi_{\rm dec}\sim
50^\circ$ with respect to the primary bar. This is facilitated by a pair of
leading shocks in the gas.   The single remaining ring is subject to a
gravitational torque from the primary bar, which acts to align the ring with
the bar. The gas, however, responds to the torque in the following way. The
total angular momentum of the gas can be assumed to consist of a circulation
along the ring with a high angular velocity $\Omega$ and of a much slower
precession $\Omega_n$ due to the torque from the bar which tumbles with
$\Omega_p$. While the latter two frequences are comparable, they are much
lower than $\Omega$. As a result, the circulation along the ring can be
treated (roughly!) as an adiabatic invariant. Therefore, the gravitational
torque changes the ring's precession rate and shape but not the internal
circulation. 

In the pre-decoupling stage, the ring becomes progressively barlike,
decreasing its axial ratio. In the low-viscosity model, this gaseous bar
resides on purely $x_2$ orbits and, therefore, responds to the primary bar
torque by speeding up its precession while being pulled backwards until it is
almost at a right angle to the bar potential valley. Higher viscosity models
have gaseous bars forming (after merging) and entering the decoupling phase
almost immediately. On the average, the gas in these bars has energies just
below that of the inner ILR, where no $x_2$ orbits exist.

The decoupling happens abruptly when a substantial fraction, $\sim1/2$, of the
gas in the bar finds itself at $E_J$ energies below the inner ILR. The absence
of $x_2$ orbits at these $E_J$ means that the bars either are unable to
settle on these orbits (Seqs.~1, 2) or lose their stable orientation along the
primary bar minor axis (Fig.~1c and Seq.~3). As a result the nuclear bars enter
librations about the main bar with smaller or larger initial departures from
the stable orientation along the main bar, {\it which is a single decisive
factor separating the partial and full decouplings}. 

The angle $\phi_{\rm dec}$ is largest for Seq.~3 --- the reason for the 
qualitatively different behavior. All three models show that the gaseous bars
in the partial or full decoupling phase experience shape changes depending
on their orientation. The bar has a much smaller eccentricity in the 4th
quadrant than in the 1st one (Figs.~2a,b,c). Such an asymmetry with respect
to the primary bar axis ensures that the resulting gravitational torques from
the primary bar are smaller in the 4th quadrant. But only for the least
viscous model (Fig.~1c), which has the maximal $\phi_{\rm dec}$, does this
make the ultimate difference. In this case the torques are unable to confine
the bar oscillation, which continues for a full swing of $2\pi$. The nuclear
bar is trapped again at $\tau\sim 211$, after many rotations with respect to
the large-scale bar. Fig.~5 shows the dramatic effect the shape change has on
the runaway. For comparison, we display the angular momentum of a hypothetical
`rigid' bar which is unable to decouple fully.

\figurenum{5}
\begin{figure}[ht!!!]
\vbox to2.7in{\rule{0pt}{2.7in}}
\includegraphics{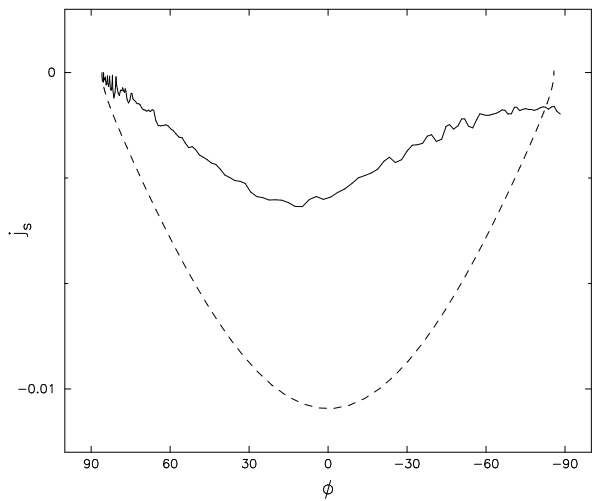}
\caption{Low-viscosity model: specific precession angular momentum $j_s$ of
gaseous bar as a function of $\phi$ during the fully decoupled phase
(starting at $\tau=150$ and $\phi_{\rm dec}=85^\circ$; first swing
only) in the reference frame of the primary bar. Positive/negative $\phi$: the
gaseous bar is leading/trailing the primary bar. {\it Solid line:} actual
$j_s$ from the numerical model (the minimum of $j_s$ is shifted from $\phi=0$
because the moment of inertia decreases with $\phi$). {\it Dashed line:} $j_s$
calculated on the assumption that the shape of the gaseous bar does not change
with $\phi$ (in this case only librations occur).}
\end{figure}       

The clear correlation between the eccentricity of the nuclear bar and
its angle with the primary bar in the decoupled phase shown by {\it all}
models can be tested observationally, e.g., by looking on the molecular rings
in CO. The maximum  in the gaseous bar intensity $\epsilon$ is achieved each
time both bars are aligned and the minimum occurs when the bars are at right
angles.  

Two additional effects should have observational consequences as a
corollary to decoupling and the periodic increase in $\epsilon$. First,
the gas will move inwards across the inner ILR on a
dynamical timescale. Shlosman et al. (1989) pointed out that the ILR(s)
present a problem for radial gas inflow because the gas can stagnate there. A
solution was suggested in the form of a global self-gravitational instability
in the nuclear ring or disk, which will generate gravitational torques in the
gas, driving it towards the nuclear region. This was confirmed by numerical
simulations (Knapen et al. 1995; Shlosman 1996). 

Recently  Sellwood \& Moore (1999) resurrected the idea that ILRs will
`choke' the gas inflow. However, as we see here, even non-self-gravitating
nuclear rings are prone to dynamical instability which drives the gas inwards,
whether in a full or partial decoupling. Second, this instability can be
accompanied by star formation along the molecular bar due to increased
dissipation in the gas, and this star formation will have a quasi-periodic
bursting character.

In summary, we have investigated the properties of non-self-gravitating gaseous
bars in barred disk galaxies and their formation from nuclear rings and  
decoupling from the underlying gravitational potential. We found that the
degree of viscosity in the gas is a crucial factor in the diverging evolution
of these  systems, namely, the low-viscosity systems are expected to spend a
substantial period of time in a fully decoupled state, with nuclear
gaseous bars having much slower pattern speeds compared to the primary bars.
The nuclear bar shape is expected to correlate with the angle between the
bars. We also find it plausible that gas compression accompanying such
decoupling, partial or full, can be associated with bursts of star formation
and with gas inflow across the ILRs towards smaller radii. Both gridless and
grid numerical codes reproduce the above evolution.

\acknowledgments

We thank Seppo Laine for his comments. This work was supported in part by NASA
grants NAG 5-3841, WKU-522762-98-6 and HST GO-08123.01-97A to I.S., which is
gratefully acknowledged.

\end{document}